\documentstyle[12pt]{article}
\begin{document}

%\hspace{100mm} UICHEP-TH/00-3
\begin{center}
\bf{ Classical solution of the wave equation }\\
\vspace{3mm}
\rm{ M. N. Sergeenko }\\
\vspace{2mm}
\it{The National Academy of Sciences of Belarus,
    Institute of Physics \\ Minsk 220072, Belarus,\\
    Homel State University, Homel 246699, Belarus \ and \\
    Department of Physics, University of Illinois at Chicago,
                 Illinois 60607, USA }
\end{center}

\begin{abstract}
The classical limit of wave quantum mechanics is analyzed. It is 
shown that the general requirements of continuity and finiteness to 
the solution $\psi(x)=Ae^{i\phi(x)}+ Be^{-i\phi(x)}$, where 
$\phi(x)=\frac 1\hbar W(x)$ and $W(x)$ is the reduced classical 
action of the physical system, result in the asymptote of the exact 
solution and general quantization condition for $W(x)$, which yields 
the exact eigenvalues of the system.\\

\noindent PACS number(s): 03.65.Sq, 03.65.Ge, 03.65.Yz \\
\end{abstract}
%\narrowtext

{\bf 1. Introduction.}
One of the fundamental principles of quantum mechanics is the Bohr's 
correspondence principle.  This principle has been used at the stage 
of creation of quantum theory and to derive the wave equation.  
Besides, this same principle results in the simplest solution of the 
Schr\"odinger equation.

There is a mathematical realization of the correspondence principle 
known as the quasiclassical approximation \cite{WKB} applicable in 
the case when the de Broglie wavelength, $\lambda =h/p$ 
($h=2\pi\hbar$), is a slowly changing function of position.  One of 
advantages of the quasiclassical approach is that we can calculate a 
tunneling effect that is beyond the usual perturbative methods.  To 
any finite order in perturbative theory, we will never see any of 
these nontrivial nonperturbative effects.  The quasiclassical 
approach also naturally leads to the conception of instantons 
\cite{Bel,Col} which have proved to be a powerful tool to probe the 
nonperturbative regime of gauge theory.  QCD instantons force us to 
re-examine the whole question of $CP$ violation \cite{tHoo}.

As well known the exact eigenvalues can be defined with the help of 
the asymptotic solution of the wave equation.  In Refs. 
\cite{SeSe,SeR} the correspondence principle has been used to derive 
the semiclassical wave equation appropriate in the quasiclassical 
region.  It was shown that solution of this equation by the standard 
WKB method yields the exact eigenvalues for {\em all} known solvable 
problems in quantum mechanics.

In this letter, we scrutinize the classical limit of the 
Schr\"odinger wave mechanics for conservative physical problems.  The 
general solution $\psi_0(x)$ of the wave equation and the 
quantization condition is written in terms of the classical action. 
The wave function ({\em wf}) in the whole region is built with the 
use of requirements of continuity and finiteness for $\psi_0(x)$ in 
the whole region.

{\bf 2. Connection formulas.}
Consider the Schr\"odinger equation for the arbitrary potential 
$V(x)$ in one dimension\footnote{ Multidimensional separable problems 
can be treated analogously.  In this case, separation should be 
performed with the help of the correspondence principle between 
classical and quantum-mechanical quantities \cite{SeQ}.},

\begin{equation}  \label{SchP}
\left(-i\hbar\frac d{dx}\right)^2\psi(x)=\left[P^2-U(x)\right]\psi(x)~,
\end{equation}
where $P^2=2mE$ and $U(x)=2mV(x)$.  As well known, in the classical 
limit the general solution of Eq. (\ref{SchP}) can be written in the 
form,

\begin{equation}  \label{psi0}
\psi_0(x) = Ae^{i\phi(x)} + Be^{-i\phi(x)},
\end{equation}
where $A$ and $B$ are the arbitrary constants and

\begin{equation}  \label{phi}
\phi(x) = \frac 1\hbar W(x)\equiv
\frac 1\hbar\int^x\sqrt{P^2-U(x)}dx
\end{equation}
is the dimensionless phase variable.  The function $W(x)$ (the 
reduced classical action of the system\footnote{For conservative 
classical system, the total action is $S_0(t,x)=-Et+W(x)$.}) 
satisfies the classical Hamilton-Jacobi equation, $(\frac{dW}{dx})^2 
=P^2-U(x)$, where $\frac{dW}{dx}\equiv p(x)$ is the generalized 
momentum.

Function (\ref{psi0}) can be treated as the ``classical'' solution of 
the Schr\"odinger equation (\ref{SchP}).  This solution has several 
remarkable features.  The most important is that the solution 
(\ref{psi0}) has no divergence at the classical turning points (TP) 
given by $E-V=0$.  Another interesting feature is that, for discrete 
spectrum, the requirements of continuity and finiteness for 
$\psi_0(x)$ in the whole region result in the asymptotic solution and 
general quantization condition for the classical action of the 
system.

Solution in quantum mechanics must be continuous finite function in 
the whole region.  To build the {\em wf} in the whole region we need 
to connect (in turning points) solution in the classically allowed 
region with solution in the classically inaccessible region.  
Consider the general solution (\ref{psi0}) in the region of the 
classical TP $x_k$.  In the classically allowed region [given by 
$E\ge V$] solution (\ref{psi0}) can be written as

\begin{equation}  \label{psiC}
\psi_0^I(\phi)=A_ke^{i(\phi-\phi_k)}+B_ke^{-i(\phi-\phi_k)}.
\end{equation}
In the classically inaccessible region [where $E<V$] the general 
solution (\ref{psi0}) is

\begin{equation}   \label{psiF}
\psi_0^{II}(\phi)=C_ke^{-\phi+\phi_k} + D_ke^{\phi-\phi_k}.
\end{equation}
The functions (\ref{psiC}) and (\ref{psiF}) must satisfy the 
continuity conditions, $\psi_0^{I}(\phi_k)=\psi_0^{II}(\phi_k)$ and 
$d[\psi_0^{I}(\phi_k)]/dx=d[\psi_0^{II}(\phi_k)]/dx$, at 
$\phi=\phi_k$ \footnote{ Note, the well-known WKB approximation 
cannot be used in the region near the TP, because when $E=V$, the 
conditions for its applicability break down.}.

Matching the functions (\ref{psiC}) and (\ref{psiF}) and their first 
derivatives at the TP $x_k$ gives

\begin{equation}
\left\{
\begin{array}{lc} \label{syst}
A_k + B_k = C_k + D_k,\\
iA_k - iB_k = -C_k + D_k.\\
\end{array}
\right.
\end{equation}
This yields

\begin{equation}
\left\{
\begin{array}{lc} \label{syspa}
A_k = \frac 1{\sqrt 2}\left(C_ke^{i\pi/4}
  + D_ke^{-i\pi/4}\right),\\
B_k = \frac 1{\sqrt 2}\left(C_ke^{-i\pi/4}
  + D_ke^{i\pi/4}\right).\\
\end{array}
\right.
\end{equation}
The connection formulas (\ref{syspa}) supply the continuous 
transition of the general solution (\ref{psiC}) into solution 
(\ref{psiF}) at the TP $x_k$.

{\bf 3. The ``classical'' {\em wf}. Quantization.}
Now we can build the ``classical'' {\em wf} for a given physical 
problem.  To build the {\em wf}, we need to choose the boundary 
conditions for the problem.  Consider first the two-turning-point 
(2TP) problem.  For the 2TP problem, the whole interval 
($-\infty,\infty$) is divided by the TP $x_1$ and $x_2$ into three 
regions, $-\infty <x<x_1$ ($I$), $x_1\le x\le x_2$ ($II$), and 
$x_2<x<\infty$ ($III$).  The classically allowed region is given by 
the interval $II$.

In the classically inaccessible regions $I$ and $III$ we choose the 
exponentially decaying solutions, i.e., $\psi_0^I(\phi) = D_1e^{\phi 
-\phi_1}$ left from the TP $x_1$ (we put $C_1=0$), and 
$\psi_0^{III}(\phi) = C_2e^{-\phi +\phi_2}$ right from the TP $x_2$ 
(here we put $D_2=0$).  Then, in the classically allowed region $II$, 
right from the TP $x_1$ we have, from (\ref{syspa}), 
$A_1=\frac{D_1}{\sqrt 2}e^{-i\pi/4}$ and $B_1=\frac{D_1}{\sqrt 
2}e^{i\pi/4}$, and solution (\ref{psiC}) takes the form,

\begin{equation}  \label{cs1}
\psi_0^{II}(\phi)=\sqrt{2}D_1\cos\left(\phi -\phi_1 -\frac\pi 4\right),
\end{equation}
and left from the TP $x_2$ [here
$A_2=\frac{C_2}{\sqrt 2}e^{i\pi/4}$ and
$B_2=\frac{C_2}{\sqrt 2}e^{-i\pi/4}$] solution (\ref{psiC}) is

\begin{equation}   \label{cs2}
\psi_0^{II}(\phi)=\sqrt{2}C_2\cos\left(\phi -\phi_2 +\frac\pi 4\right),
\end{equation}
where $\phi_1=\phi(x_1)$ and $\phi_2=\phi(x_2)$.
We see that the superposition of two plane waves (\ref{psi0}) in the 
phase space results in the standing wave given by Eqs. (\ref{cs1}) 
and (\ref{cs2}).

Functions (\ref{cs1}) and (\ref{cs2}) should coincide in each point 
of the interval $[x_1,x_2]$.  Putting $\phi=\phi_2$ we have, from 
Eqs. (\ref{cs1}) and (\ref{cs2}),

\begin{equation}  \label{cs12}
D_1\cos\left(\phi_2-\phi_1 -\frac\pi 4\right)=C_2\cos\frac\pi 4~.
\end{equation}
This equation is valid if

\begin{equation}  \label{eq}
\phi_2 - \phi_1 -\frac\pi 4 = \frac\pi 4 + \pi n ,~~
n = 0,1,2,\ldots
\end{equation}
and $D_1=(-1)^nC_2$.
Equation (\ref{eq}) is the condition of the existence of continuous 
finite solution in the whole region.  This condition being, at the 
same time, quantization condition.  Taking into account the notation 
(\ref{phi}), we have, from Eq. (\ref{eq}),

\begin{equation}  \label{qc2}
\int_{x_1}^{x_2}\sqrt{P^2-U(x)}dx = \pi\hbar\left(n +\frac 12\right).
\end{equation}
Condition (\ref{qc2}) solves the 2TP eigenvalue problem given by Eq. 
(\ref{SchP}).  Note, we have obtained Eq. (\ref{qc2}) by product from 
requirements of a smooth transition from oscillating solution 
(\ref{psiC}) to the exponentially decaying solutions in the 
classically inaccessible regions.  The quantization condition 
(\ref{qc2}) reproduces the exact eigenvalues for {\em all} known 2TP 
problems in quantum mechanics (see Refs. \cite{SeSe,SeR}).

Combining the above results, we can write the finite continuous 
solution in the whole region (the ``classical'' wave function),

\begin{equation}
\psi_0[\phi(x)] = C\left\{
\begin{array}{lc}  \label{osol}
\frac 1{\sqrt 2}e^{\phi(x) -\phi_1}, & x<x_1,\\
\cos[\phi(x) -\phi_1 -\frac\pi 4], & x_1\le x\le x_2,\\
\frac{(-1)^n}{\sqrt 2}e^{-\phi(x) +\phi_2}, & x>x_2.
\end{array}
\right.
\end{equation}
Oscillating part of solution (\ref{osol}) corresponds to the main 
term of the asymptotic series in theory of the second-order 
differential equations.  In quantum mechanics, the oscillating part 
of Eq. (\ref{osol}) gives the asymptote of the exact solution of the 
Schr\"odinger equation.

An important consequence of the ``classical'' solution (\ref{psi0}) 
is conservation not only energy $E$, but also momentum.  In Eq. 
(\ref{SchP}), $P^2=2mE$ is a constant and, as $E$, it can take only 
discrete values.  Substituting (\ref{psi0}) into Eq. (\ref{SchP}), we 
have

\begin{equation}   \label{equ0}
\left(\frac{dW}{dx}\right)^2
-i\hbar\frac d{dx}\left(\frac{dW}{dx}\right)=P^2-U(x).
\end{equation}
So far, as $(\frac{dW}{dx})^2= P^2-U(x)$, we obtain the constraint,

\begin{equation}  \label{cotr}
\hbar\frac d{dx}\left(\frac{dW}{dx}\right)=0.
\end{equation}

The constant $\hbar$ is a small value, which is used as the expansion 
parameter in the quasiclassical approximation.  The constraint 
(\ref{cotr}) can be achieved not only by taking the limit 
$\hbar\rightarrow 0$ (that leads to the classical mechanics), but 
also by assuming that\footnote{ The well known WKB approximation is 
based on the condition ${W^\prime}^2\gg\left|\hbar 
W^{\prime\prime}\right|$ that implies that the momentum 
$p(x)=W^\prime(x)$ is large enough.}

\begin{equation}  \label{cotr1}
\frac{dW}{dx}\simeq {\rm const}.
\end{equation}
This implies the adiabatically slow alteration of the momentum 
$p(x)\equiv W^\prime(x)$.  In this case, $\hbar W^{\prime\prime}(x)$ 
is a small value of the second order and can be neglected.  Note, the 
constraint (\ref{cotr1}) supplies the Hermiticity of the operator 
$\hat p^2=(-i\hbar\frac d{dx})^2$ in Eq. (\ref{SchP}) \cite{SeSe}; in 
order for the operator $\hat p^2$ to be Hermitian the expression 
${W^\prime}^2-i\hbar W^{\prime\prime}$ should be real.

Underline, the potential $V(x)$ in Eq. (\ref{SchP}) is not a 
constant.  The constraint (\ref{cotr1}) is the requirement to the 
{\em final} solution we build from the general expression 
(\ref{psi0}); it is valid only for certain values of 
$P_n^2=\left<p^2(x)\right>$, which correspond to the discrete values 
of the action $W(x)$ given by Eq. (\ref{qc2}).  This is a specific 
requirement for the allowed motions in quantum mechanics: for the 
conservative quantum-mechanical systems, the particle momentum in 
stationary states is the {\em integral of motion}.

According to Bohr's postulates, 1) electrons in atomic orbits are in 
stationary states, i.e. do not radiate, despite their acceleration, 
and 2) electrons can take discontinuous transitions from one allowed 
orbit to another.  However, the postulate is not an answer to the 
question: why the accelerated electrons do not radiate?

Acceleration means change in momentum. As we have shown above, 
particles in the stationary states are {\em not} accelerated, because 
the momentum is the integral of motion.  The fact the momentum is a 
constant value means the electrons in stationary states move like 
free particles.  In these states, the momentum eigenvalues and the 
energy eigenvalues are connected with the help of the equality 
$P_n^2=2mE_n$ for free particles.  Such kind of wave motion is 
described by the standing waves.

The oscillating part of the {\em wf} (\ref{osol}) is in agreement 
with the asymptote of the corresponding exact solution of Eq. 
(\ref{SchP}) and can be written in the form of a standing wave.  
Integrating (\ref{cotr1}), we obtain $W(x,n)= P_nx+{\rm const}$ that 
gives, for the function $\psi_0^{II}(x)$,

\begin{equation}  \label{fn0}
\psi_0^{II}(x) = C_n\cos\left(k_nx +\frac\pi 2 n\right),
\end{equation}
where $k_n=P_n/\hbar$.
The normalization coefficient, $C_n=\sqrt{2k_n/[\pi(n+\frac 12)+1]}$, 
is calculated from the normalization condition $\int_{-\infty}^\infty 
\left|\psi_0(x)\right|^2=1$.

In Eq. (\ref{fn0}), we have took into account the fact that, in the 
stationary states, the phase-space integral (\ref{phi}) at the TP 
$x_1$ and $x_2$ is $\phi_1=-\frac\pi 2(n+\frac 12)$ and 
$\phi_2=\frac\pi 2(n+\frac 12)$, respectively, [so that 
$\phi_2-\phi_1=\pi(n+\frac 12)$] \cite{SeSe,SeQ}, i.e. it depends on 
quantum number and does not depend on the form of the potential.  
This form of $\phi_1$ and $\phi_2$ guaranties that the eigenfunctions 
are necessarily either symmetrical ($n=0,2,4,\ldots$) or 
antisymmetrical ($n=1,3,5,\ldots$).  Solution (\ref{fn0}) describes 
free motion of a particle-wave in the enclosure (the enclosure being 
the interaction potential).  Therefore, in bound state region, the 
interaction of a particle-wave with the potential reduces to 
reflection of the wave by the walls of the potential.

The ``classical'' solution (\ref{psi0}) is general for all types of 
problems and allows to solve multi-turning-point problems ($M$TP, 
$M>2$), i.e. a class of the ``insoluble'' problems, which cannot be 
solved by standard methods.  In the complex plane, the 2TP problem 
has one cut between turning points $x_1$ and $x_2$, and the 
phase-space integral (\ref{qc2}) can be written as the contour 
integral about the cut.  The $M$TP problems contain (in general case) 
bound state regions and the potential barriers, i.e. several cuts.  
The corresponding contour should enclose all cuts.  The ``classical'' 
{\em wf} in the whole region can be built similar to the 2TP problem 
with the help of the same connection formulas (\ref{syspa}).

Let the problem has $\nu$ cuts. Then the integral about the contour 
$C$ can be written as a sum of contour integrals about each of the 
cut, where each term of the sum is the 2TP problem.  Then, the $2\nu$ 
TP quantization condition can be written as \cite{SeR,SeQ}

\begin{equation}  \label{genC}
\oint_C\sqrt{P^2-U(z)}dz = 2\pi\hbar\left(N+\frac\mu 4\right),
\end{equation}
where $N=\sum_{k=1}^{\nu}n_k$ is the total number of zeroes of the 
{\em wf} on the $\nu$ cuts and $\mu=2\nu$ is the number of turning 
points, i.e. number of reflections of the {\em wf} by the walls of 
the potential (Maslov's index \cite{MaslFed}).

{\bf 4. Relativistic Cornell problem, $V(r)=-\frac{\tilde\alpha} r
+\kappa r$.}
To demonstrate efficiency of the ``classical'' solution, let us 
obtain the exact energy eigenvalues for the famous funnel type 
potential (Cornell potential) \cite{Eich}.  This potential is one of 
a special interest in high-energy hadron physics, quarkonium physics, 
and quark potential models.  Its parameters are directly related to 
basic physical quantities of hadrons: the universal Regge slope 
$\alpha '\simeq 0.9\,($GeV$/c)^{-2}$ of light mesons and 
one-gluon-exchange coupling strength, $\alpha_s$, of heavy quarkonia.

In relativistic theory, the Cornell potential represents the 
so-called ``insoluble'' 4TP problem. The ``classical'' solution and 
quantization condition derived above give the asymptote of the exact 
solution and yield the exact energy spectrum for the potential.  A 
series of examples has been considered elsewhere \cite{SeSe,SeR,SeQ}, 
where the exact energy spectrum of quantum systems was reproduced 
from quasiclassical solution of the semiclassical wave equation 
\cite{SeSe,SeR}.

For the system of two particles of equal masses, the relativistic 
radial semiclassical wave equation for the Cornell potential is 
($\hbar=c=1$) \cite{SeR}

\begin{equation}   \label{qcPot}
\left(-i\frac d{dr}\right)^2\tilde R(r) =\left[\frac{E^2}4 -
\left(m -\frac{\tilde\alpha}r +\kappa r\right)^2 -\frac{(l+\frac
12)^2}{r^2}\right]\tilde R(r),
\end{equation}
where $\tilde\alpha =\frac 43\alpha_s$.  The quantization condition 
(\ref{genC}) appropriate to Eq. (\ref{qcPot}) is $I =\oint_C[\frac 
14E^2 -(m -\tilde\alpha/r+\kappa r)^2 -(l+\frac 12)^2/r^2]^{\frac 
12}dr$ $=4\pi\left(n_r+\frac 12\right)$.  To calculate this integral, 
we use the method of stereographical projection that gives $I=I_0 
+I_{\infty}$, where $I_0=-2\pi\Lambda$ and $\Lambda=[(l+\frac 12)^2 
+\tilde\alpha^2]^{\frac 12}$.  The integral $I_{\infty}$ is 
calculated with the help of the replacement of variable, $z=\frac 
1r$, that gives $I_{\infty} =2\pi[E^2/8\kappa +\tilde\alpha]$. 
Therefore, for $E_n^2$, we obtain

\begin{equation}  \label{EnExac}
E_n^2=8\kappa\left[2\left(n_r+\frac 12\right) +\Lambda
-\tilde\alpha\right].
\end{equation}
This is the exact result for the Cornell potential.

It is an experimental fact that the dependence $E_n^2(l)$ is linear 
for light mesons.  At present, the best way to reproduce the 
experimental masses of particles is to rescale the entire spectrum 
for the linear potential, $E_n^2=8\kappa(2n_r+l+\frac 32)$, assuming 
that the masses of the mesons are expressed by the relation 
$M_n^2=E_n^2-C^2$, where $C$ is an additional free parameter.  At the 
same time, Eq. (\ref{EnExac}) does not require any additional free 
parameter. We obtain the necessary shift with the help of the term 
$-8\kappa\tilde\alpha$ which is the result of interference of the 
Coulomb and linear terms of the interquark potential.  The formula 
describes the light meson trajectories with the accuracy $\simeq 
1\%$.

{\bf 5. Discussion and Conclusion}.
In the classical limit $\hbar\rightarrow 0$, the quantum-mechanical 
action $S(t,x)$ reduces to the classical action $S_0(t,x)$, i.e. 
$S(t,x)=S_0(t,x)$.  Then, for conservative systems, we can write 
$S(t,x)=-Et+W(x)$, where $W(x)$ is the reduced classical action, and 
the general solution $\psi_0(x)$ of the wave equation can be written 
in the form of superposition of two waves, $\exp[\pm iW(x)/\hbar]$.

Using the general requirements of continuity and finiteness for 
$\psi_0(x)$ and $\psi_0^\prime(x)$, we have derived simple connection 
formulas that allowed us to build the ``classical'' {\em wf} in the 
whole region and the corresponding quantization condition for the 
classical action.  If well-known quasiclassical approximation is 
applicable for a distance from the turning point satisfying the 
condition $\left|x-x_0\right|\gg\lambda/4\pi$ and potentials 
satisfying the semiclassical condition, $m\hbar V^\prime\ll 
2m[E-V]^{3/2}$, the ``classical'' solution considered in this Letter 
is applicable in the whole region and for any separable potentials.

The final solution has written in terms of elementary functions and 
corresponds to the main term of the asymptotic series in the theory 
of the second-order differential equations.  We have observed that, 
for the conservative systems, not only energy, but also momentum is 
the integral of motion.  This means that particles in stationary 
states move like free particles-waves in enclosures.  This has 
allowed us to write the ``classical'' solution in the form of the 
standing wave, which describes free finite motion of particles-waves 
in enclosures.

There is a simple connection of the ``classical'' solution considered 
here with the Feynman path integrals: this solution corresponds to 
the path of ``minimum'' action.  The ``classical'' solution obtained 
corresponds to the classical path.

{\it Acknowledgements}.
The author thanks Prof. U. P. Sukhatme for kind invitation to visit 
the University of Illinois at Chicago and Prof. A. A. Bogush for 
support and constant interest to this work.\\ This work was supported 
in part by the Belarusian Fund for Fundamental Research.

\end{document}